\documentclass{aastex63}

\begin{document}

\title{Vintage NPOI: New and Updated Angular Diameters for 145 Stars}

\author{Ellyn K. Baines}
\affil{Naval Research Laboratory, Remote Sensing Division, 4555 Overlook Ave SW, Washington, DC 20375, USA}
\email{ellyn.k.baines.civ@us.navy.mil}

\author{James H. Clark III}
\affil{Naval Research Laboratory, Remote Sensing Division, 4555 Overlook Ave SW, Washington, DC 20375, USA}

\author{Bradley I. Kingsley}
\affil{Naval Research Laboratory, Remote Sensing Division, 4555 Overlook Ave SW, Washington, DC 20375, USA}

\author{Henrique R. Schmitt}
\affil{Naval Research Laboratory, Remote Sensing Division, 4555 Overlook Ave SW, Washington, DC 20375, USA}

\author{Jordan M. Stone}
\affil{Naval Research Laboratory, Remote Sensing Division, 4555 Overlook Ave SW, Washington, DC 20375, USA}

\begin{abstract}

We present new or updated angular diameters, physical radii, and effective temperatures for 145 stars from the Navy Precision Optical Interferometer data archive. We used data from 1996 to late-2021, and we describe the differences between early and late data, which hinge upon an update of the beam combiner in 2002. We came across several sub-categories of stars of interest: 13 of our stars are promising targets for the Habitable World Observatory and therefore require as much study as possible, and 14 more are asteroseismic targets and have stellar masses after we combined our radii and effective temperatures with frequencies of maximum oscillation power values from the literature. In addition to this, many of the stars here show measurements to the first null in the visibility curve and beyond, which is the gateway to determining second-order effects such as direct measurements of limb-darkening. Finally, we consider the stars in the larger context of previous NPOI measurements and find the majority (75$\%$) of the angular diameters in the overall NPOI sample have uncertainties of 2$\%$ or less.

\end{abstract}


% ------------------------------------------------

\section{Introduction} 

The number of stellar surveys has boomed in recent years, as have the quality of the resulting data, with missions such as Gaia \citep{2016AandA...595A...1G}, Galactic Archaeology with the High Efficiency and Resolution Multi-Element Spectrograph \citep[GALAH, ][]{2015MNRAS.449.2604D}, Apache Point Observatory Galactic Evolution Experiment \citep[APOGEE, ][]{2008AN....329.1018A}, and so on. In order to get the highest scientific return from these surveys, we need to understand a sample of benchmark stars as completely as possible so the stellar models applied to the surveys can be properly scaled \citep{2022AandA...658A..48K}. This is accomplished by providing as many directly obtained measurements as we can, including effective temperature ($T_{\rm eff}$). Most often this is obtained using spectroscopy, which is an inherently indirect method to get $T_{\rm eff}$. Optical interferometry is the solution to the problem, directly measuring a star's angular diameter and then calculating its $T_{\rm eff}$ with the weakly model-dependent parameters of bolometric flux and limb-darkening coefficients.

Since 2018, we have been publishing batches of stellar diameters from the Navy Precision Optical Interferometer's (NPOI) data archive. This has resulted in nearly 180 measurements spanning three publications \citep[][and when we began, we did not realize there would be anything past the first paper to add!]{2018AJ....155...30B, 2021AJ....162..198B, 2023AJ....166..268B}. In addition to that, we recently published a catalog of 588 stellar objects with NPOI data including the year(s) of observation(s) and how many scans were obtained for each target \citep{10.1117/12.3027444} in the hope of fostering collaborations between groups and making the data more easily shared with the astronomical community. 

During this effort, we realized that a good number of the stars either included early NPOI data (from 1996 to mid-2002) or later NPOI data (mid-2002 on), with the vast majority of the published data being from the latter. We began investigating the earlier data to see how it could contribute to published and new angular diameters. The result is a sample of 145 stars: 62 stars with new diameters and 83 stars with updated diameters that include both early and late NPOI data. 

\begin{figure}[h]
\includegraphics[width=0.85\textwidth]{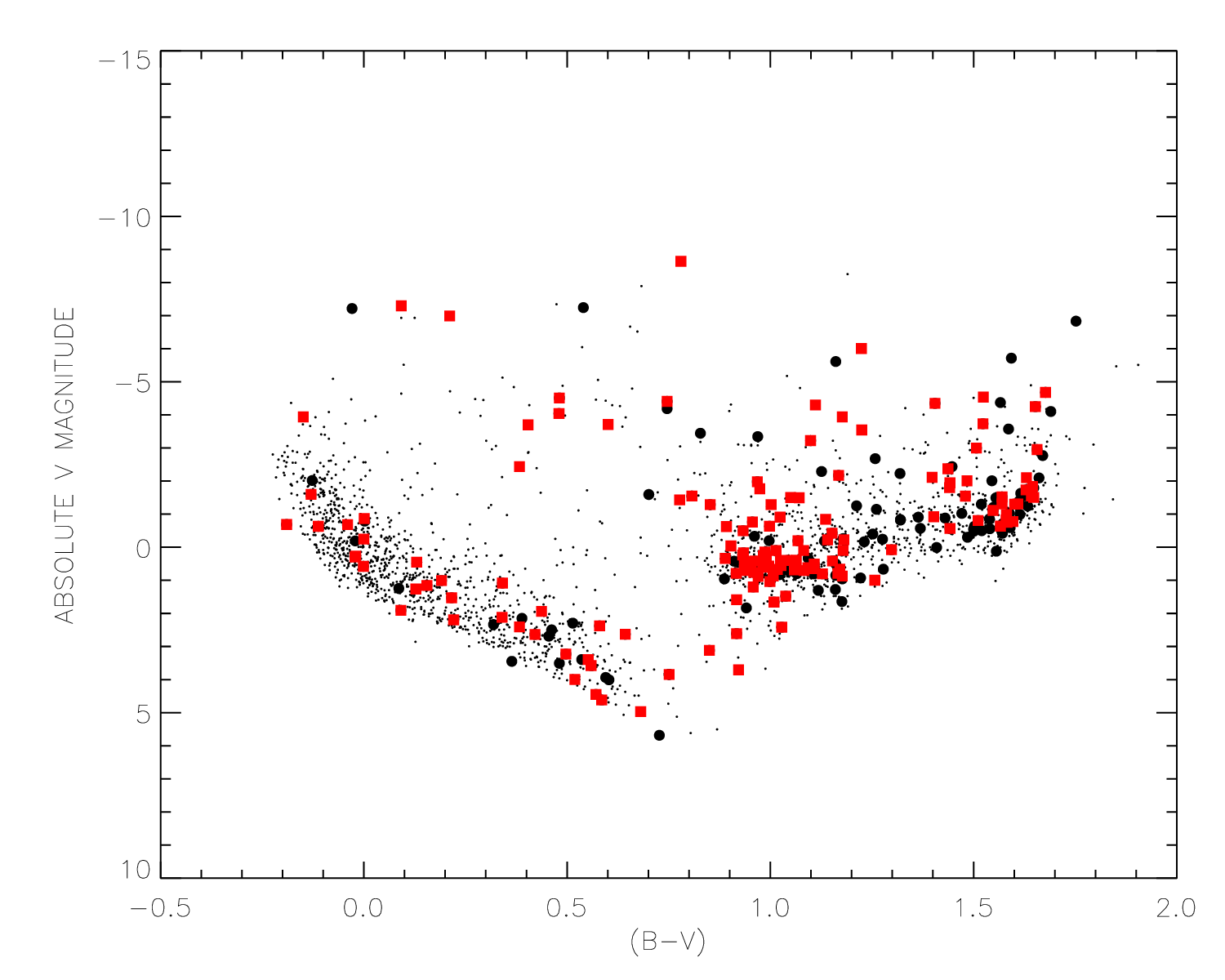}
\caption{A color-magnitude diagram of the stars presented here (red squares), past NPOI targets (large black circles), and targets from JSDC (small gray points) that are observable with the NPOI (decl. $>$ $-$10$^{\circ}$, $V$ $<$ 6.0).}
  \label{color_mag}
\end{figure}

\begin{figure}[h]
\includegraphics[width=0.85\textwidth]{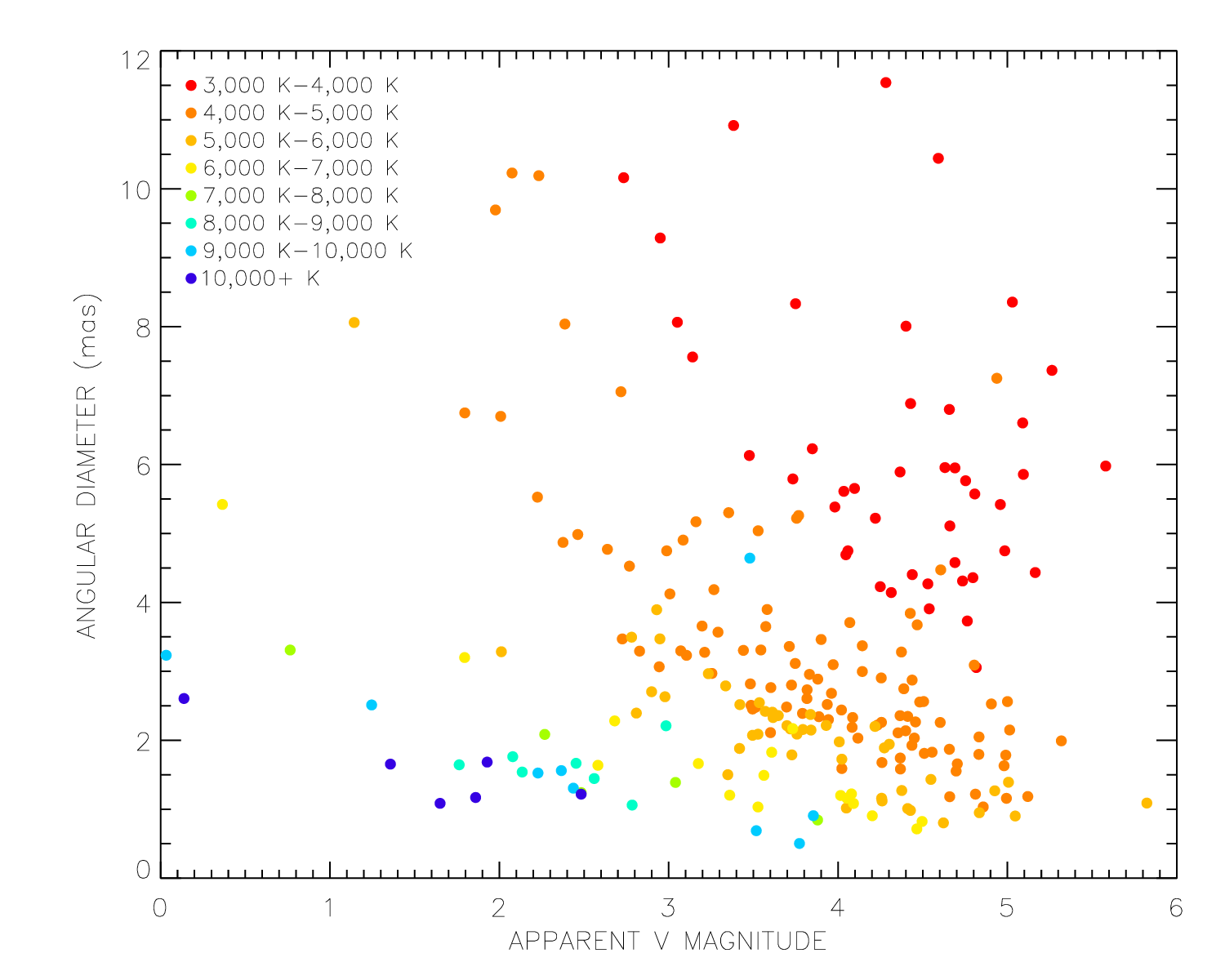}
\caption{We plot angular diameters versus apparent $V$ magnitudes with the temperature index listed in the upper left corner.}
  \label{vdt}
\end{figure}

Figure \ref{color_mag} puts the stars observed here in the framework of a color-magnitude diagram of stars observable with the NPOI. In this plot, we compare our sample to JMMC Stellar Diameters Catalog \citep[JSDC, ][]{2014ASPC..485..223B} that fall within the NPOI's observable range with respect to declination and brightness. We cross-correlated the JSDC stars with parallaxes from \citet[][and we note we would have used Gaia values but many stars in the JSDC were missing parallaxes from Gaia]{2007AandA...474..653V}. These add up to nearly 2180 stars, and they are shown as small grey points on the Figure \ref{color_mag}. As a side note, the NPOI has measured $\sim 19 \%$ (212 of 1146) of the JSDC stars in Figure \ref{color_mag} that are larger than 0.8 mas, and this does not include the 22 stars measured by the NPOI that are not in the JSDC. Figure \ref{vdt} shows another representation of all NPOI angular diameters as they relate to apparent $V$ magnitude and $T_{\rm eff}$, where the temperatures are color coded.

The paper is organized as follows: Section \ref{sec:npoi} reviews the NPOI and the data collection and reduction process, Section \ref{sec:obsvis} describes the observations and defines interferometric visibility, Section \ref{sec:randt} summarizes the process of determining stellar radii and $T_{\rm eff}$, Section \ref{sec:notes} considers various aspects about individual stars of interest, Section \ref{sec:disc} is the discussion and conclusion, where we merge the sample presented here with previously published NPOI angular diameters with respect to potential Habitable World Observatory targets, stellar oscillators, and zero-crossing stars.

% ------------------------------------------------

\section{The NPOI and Data Analysis}
\label{sec:npoi}

The NPOI is an optical interferometer on Anderson Mesa, near Flagstaff, AZ, and has been in operation since 1996 \citep{1998ApJ...496..550A}. For the handful of years from 1996 to mid-2002, the original version of the ``Classic'' beam combiner was used. It recorded data on one baseline\footnote{The baseline is the distance between two siderostats or telescopes. Other optical/infrared interferometers use telescopes but the NPOI uses siderostats, which are flat mirrors.} per spectrograph, of which there were three, and it observed over 32 spectral channels from 430 to 860 nm. Each 2-ms data frame was split into eight temporal bins, and each scan was 90 s long. The data reduction for the data for these early years followed the procedures described in \citet{1998AJ....116.2536H}.

Due to the limited amount of data storage available in the fringe tracker and the limited data transfer bandwidth between the fringe tracker and the data storage system, starting in mid-2002 the NPOI system changed to recording data from 16 channels (550 to 860 nm) and two spectrographs, using 30-s scans. Although this approach reduced the wavelength coverage of the instrument, in practice this did not affect a large number of targets, given the low sensitivity of the system below 550 nm. The slight loss in wavelength coverage is more than compensated for by the larger number of baselines that can be simultaneously observed. NPOI implemented the combination of light from up to four siderostats per spectrograph, corresponding to six baselines per spectrograph. This procedure allowed for the simultaneous recording of all 15 baselines observable with six siderostats (when three spectrographs were in use), but required the use of 64 temporal bins for each 2-ms data frame in order to be able to decode the measurements from the multiple baselines. 

The data collection and reduction has since remained the same, using the updated version of the ``Classic'' beam combiner \citep{2003AJ....125.2630H, 2016ApJS..227....4H}. During this time, each observation consisted of a pair of scans: one 30-s coherent (on the fringe) scan where the fringe contrast was measured every 2 ms, and one incoherent (off the fringe) scan used to estimate the additive bias affecting fringe measurements. These measurements used the data reduction package \emph{OYSTER} that was developed by C. A. Hummel\footnote{www.eso.org/$\sim$chummel/oyster/oyster.html} and automatically edits data as described in \citet{2003AJ....125.2630H}.

Over the last almost 30 years, the NPOI has observed hundreds of stars, and here we present measurements for 145 of them. Table \ref{baselines} shows the range of baselines used in this study, which go from a mere 7.2 m to 79.4 m. Table \ref{general_properties} consists of a list of the stars measured here with their identifiers, spectral type from SIMBAD, $V$ magnitude from \citet{Mermilliod}, parallax from Gaia's various data releases or Hipparcos \citep{2007AandA...474..653V}, rotational velocity ($v \sin i$) from \citet{{1964cbs..book.....H}}, and whether or not they were previously published by the authors of this paper. Table \ref{table_refs} is a list of the reference abbreviations used in all of the tables, and Table \ref{observations} shows a portion of the observing log.  

\clearpage

% ------------------------------------------------

\section{Observations and Visibilities}
\label{sec:obsvis}

The data set presented here is by far the largest every published in one place from the NPOI, totaling very nearly 360,000 data points. The dates of observations range from May, 1996 to when the NPOI temporarily shut down after October, 2021. All of the scientific targets were observed alternately with calibrator stars, which are chosen to be small, i.e., significantly less than the formal resolution of the baseline used for a given observation, and symmetrical. This means the scientific target's angular diameter measurement would be only weakly dependent on the calibrator star's angular diameter determination \citep{2018AJ....155...30B}. 

\subsection{Calibrator Stars}

We needed to begin with the calibrator star's diameter, so we fit each calibrator's spectral energy distribution (SED) to published \emph{UBVRIJHK} photometry. We used plane-parallel model atmospheres from \citet{2003IAUS..210P.A20C} based on $T_{\rm eff}$, surface gravity (log~$g$), and $E(B-V)$ also from the literature. The stellar models were fit to observed photometry after converting magnitudes to fluxes using \citet{1996AJ....112..307C} for $UBVRI$ and \citet{2003AJ....126.1090C} for $JHK$. Table \ref{calibrators} provides the information used in the SED fitting: \emph{UBVRIJHK} magnitudes and their sources, $T_{\rm eff}$, log $g$, and $E(B-V)$ with their respective sources, and the resulting diameter estimate. The errors on the diameter estimates are 5$\%$, which is what \emph{OYSTER} automatically assigns.

When observing, we made two basic assumptions about the scientific targets: that they are effectively single, and that they do not rotate rapidly enough to be oblate and experience gravity darkening \citep{2012AandARv..20...51V}. A good number of the targets discussed here do indeed have stellar companions, but in the majority of cases, those companions are well outside the NPOI's detection sensitivity. \citet{2016ApJS..227....4H} demonstrated that the NPOI can detect binaries with separations from 3 to 860 mas with magnitude differences ($\Delta m_V$) $<$ 3.0 for most binary systems, and up to 3.5 when the component spectral types differ by less than two. Section \ref{sec:notes} includes details on the stars that may fall within these windows.

\subsection{Angular Diameter Measurements}

Generally, interferometric measurements are known as visibility squared ($V^2$), which in essence goes from 0, where the star is completely resolved, to 1, where the star is completely unresolved. We first fit a uniform disk to the $V^2$: $V^2 = [2 J_1(x) / x]^2$, where $J_1$ is the Bessel function of the first order, and $x = \pi B \theta_{\rm UD} \lambda^{-1}$, where $B$ is the projected baseline toward the star's position, $\theta_{\rm UD}$ is the apparent uniform disk angular diameter of the star, and $\lambda$ is the effective wavelength of the observation \citep{1992ARAandA..30..457S}.  We then applied a limb-darkening coefficient ($\mu_{\rm LD}$) because limb darkening is a much more realistic vision of the star than a uniformly illuminated disk. 

We used $T_{\rm eff}$, log $g$, and metallicity ([Fe/H]) values from the literature with a microturbulent velocity of 2 km s$^{\rm -1}$ to obtain $\mu_{\rm LD}$ from \citet{2011AandA...529A..75C}. We used the ATLAS model in the \emph{R}-band, since that waveband most closely matched the central wavelength of the NPOI's bandpass. Because the $\mu_{\rm LD}$ are presented in tabular form, we interpolated between values when required.

For each limb-darkened angular diameter ($\theta_{\rm LD}$) fit, the uncertainties were derived via the reduced $\chi^2$ minimization method \citep{1992nrca.book.....P, 2003psa..book.....W}: the diameter fit with the lowest $\chi^2$ was determined and the corresponding diameter was the final $\theta_{\rm LD}$ for the star. The uncertainties were calculated by finding the diameter at $\chi^2 + 1$ on either side of the minimum $\chi^2$ and determining the difference between the $\chi^2$ diameter and $\chi^2 +1$ diameter.

Figure \ref{diam_fit} shows of an angular diameter fit as an example, and fits are included for all the stars in the online version of the \emph{Astronomical Journal}. Note that the y-axis is $V^2$ as expected, and  x-axis is spatial frequency, which folds the baseline and the wavelength into one value (spatial frequency $u = B/\lambda$). Table \ref{inf_results} lists each star's $\theta_{\rm UD}$, the $T_{\rm eff}$, log$g$, and metallicity ([Fe/H]) used to calculate the initial $\mu_{\rm LD}$ and its $\theta_{\rm LD}$, and the final $\mu_{\rm LD}$ and $\theta_{\rm LD}$ (see the next section of explanation of the initial and final $\mu_{\rm LD}$). The maximum spatial frequency of the data set for each star is included in the table, as well as the number of calibrated data points used in the angular diameter fit. 

% ------------------------------------------------

\section{Stellar Radius and Effective Temperature}
\label{sec:randt}

To convert the angular diameters to stellar radii, we combined our new and updated angular diameter measurements with parallaxes from the literature to calculate the physical size of the star $R$. Uncertainties for stellar radii ($\sigma_{\rm R}$) were propagated from uncertainties in the parallax ($\sigma_{\rm par}$) and angular diameter measurements ($\sigma_{\rm LD}$). Generally speaking, $\sigma_{\rm LD}$ contributed an average of 1.7$\%$ to $\sigma_{\rm R}$ versus an average of 2.4$\%$ from $\sigma_{\rm par}$. The minimum and maximum $\sigma_{\rm R}$ were 0.03$\%$ and 21$\%$ for $\sigma_{\rm par}$, respectively, and 0.1$\%$ and 46$\%$ for $\sigma_{\rm LD}$.

We obtained luminosity $L$ values from the literature, and calculated bolometric flux $F_{\rm BOL}$ by inverting $L = 4 \pi d^2 F_{\rm BOL}$, where $d$ is the star's distance. We then combined $F_{\rm BOL}$ and $\theta_{\rm LD}$ to determine each star's $T_{\rm eff}$ using Stefan-Boltzmann's equation:
\begin{equation}
F_{\rm BOL} = {1 \over 4} \theta_{\rm LD}^2 \sigma T_{\rm eff}^4,
\end{equation}
where $\sigma$ is the Stefan-Boltzmann constant and $\theta_{\rm LD}$ is in radians \citep{1999AJ....117..521V, 2014MNRAS.438.2413V}. The resulting $R$, $F_{\rm BOL}$, and $T_{\rm eff}$ are listed in Table \ref{derived_results}. Uncertainties in $T_{\rm eff}$ were determined by propagating the uncertainties from $F_{\rm BOL}$ (and therefore $L$ and $\sigma_{\rm par}$) and $\sigma_{\rm LD}$.

We performed an iterative process to arrive at the final $\mu_{\rm LD}$, $\theta_{\rm LD}$, and $T_{\rm eff}$ because $T_{\rm eff}$ is used to select $\mu_{\rm LD}$. We calculated the initial $\theta_{\rm LD}$ using $T_{\rm eff}$ from the sources listed in Table \ref{inf_results} to select $\mu_{\rm LD}$, and then used our new $T_{\rm eff}$ to update $\mu_{\rm LD}$, recalculate $\theta_{\rm LD}$, and then recalculate $T_{\rm eff}$. For 43 of our 145 stars, $\mu_{\rm LD}$ did not change at all during the process, and the rest converged after just one iteration. Overall, $\mu_{\rm LD}$ did not change much, with an average of 0.02 and a maximum of 0.16. This produced a change $\theta_{\rm LD}$ by an average of 0.3$\%$ (0.010 mas) and at most 1.8$\%$ (0.142 mas). The effect was similarly small for $T_{\rm eff}$ with an average change of 0.2$\%$ (9 K) and the largest change of 1.3$\%$ (145 K, which was for one of the hottest stars in our sample HD 87901/$\alpha$ Leo/Regulus, $T_{\rm eff} >$ 11,000 K). 

% ------------------------------------------------

\section{Notes on Individual Stars Published Here}
\label{sec:notes}

A good number of our stars have entries in the Washington Visual Double Star Catalog \citep[WDS,][]{2001AJ....122.3466M} but it is mostly not an issue for our observations. This is most often because the magnitude difference ($\Delta m_V$) is too large for us to detect the companion, sometimes because the binary is at too wide of a separation ($a$), and occasionally both (see Table \ref{wds}). Stars requiring further discussion with respect to the WDS and other interesting factors are included below. The stars with asterisks should be used especially carefully, due to potential bias affecting the diameters as described.

\begin{itemize}

% ------------------------------------------------

\item \emph{HD 5394/$\gamma$ Cas}: The SIMBAD Astronomical Database \citep{2000AandAS..143....9W} labels this as a high-mass X-ray binary, which barely begins to explore why this star is intriguing. $\gamma$ Cas is a complex and intensively studied object with an H-$\alpha$ envelope \citep{2023AJ....165..117M}, reflecting nebulae \citep{2024MNRAS.529.1680E}, a long history of multi-wavelength variability that has come and gone over the last decades \citep{2021ApJ...915...13S}, and an expanding cavity driven by stellar wind and binarity \citep{2021ApJ...922..183C}. It is also the prototypical classical Be star, discovered in 1866 by Angelo Secchi from its H-$\alpha$ spectral line, and one of the first emission-line stars ever discovered \citep{1867sspd.book.....S, 2003PASP..115.1153P}.

In addition to all this, $\gamma$ Cas has various other quirks related to its multiplicity: \citet{2021AJ....161..144T} describes it as a quadruple system consisting of a spectroscopic pair ($P$=203 d), a 60-year astrometric subsystem, and a faint physical companion at 2.05 arcsec. \citet{2024ApJ...967....8P} used adaptive optics to image the star and found two companions, one of which is bright and ``physically associated'' but ``not bound'' to $\gamma$ Cas, and one of which is faint and proper motion studies show it is likely unrelated to the primary star. \citet{2021ApJS..257...69H} did not detect the companion with the NPOI, and \citet{2024ApJ...962...70K} pointed to a massive white dwarf as the companion. Finally, the WDS shows the closest pair AB ($a \sim 2$ arcsec) to have $\Delta m_V =$ 8.7 mag.

% ------------------------------------------------

\item \emph{HD 7087/$\chi$ Psc}: This is a lithium-rich giant, which is a rarity \citep{2015AJ....150..123R}. These kinds of stars represent less than 1$\%$ of stars \citep{2014ApJ...785...94L}.   

% ------------------------------------------------

\item \emph{HD 8890/$\alpha$ UMi}: Best known as Polaris or the North Star, this classical Cepheid variable has been widely studied, and it was even part of a study to directly check the speed of light's constancy outside our solar system for the first time \citep{1981PASP...93..777G}. Polaris' multiplicity specifically has a long and storied history, including the first detection from \citet{1937PASP...49..202W} using ``an interferometer,'' per Wilson's paper, and ``an eyepiece interferometer'' per the WDS. Much later, the Aa,Ab pair was resolved using the Advanced Camera for Surveys on HST by \citet{2008AJ....136.1137E} with $\Delta m_V = 5.4$ mag at 225.5 nm and an estimated 7.2 mag in the visible. Note that this is at odds with the WDS catalog's $\Delta m_V$ of 2.0 mag. More recently, \citet{2024ApJ...971..190E} used the CHARA Array to determine the orbit and dynamical masses, and found the Cepheid is more luminous than predicted by evolutionary tracks. With respect to Polaris' Cepheid nature, our observations only span a couple of months, so we would not expect to detect the pulsations.

% ------------------------------------------------

\item \emph{HD 11353$^\ast$/$\zeta$ Cet}: SIMBAD lists this star as a spectroscopic binary, though the WDS does not include the $\Delta m_V$ for the inner-most Aa,Ab pair. The period ($P$) is long (1652 d), and $\Delta m_V$ for the outer pair AB is 6.4 mag. We present the star as single here, but note that depending on the brightness, the close companion could impact the visibility measurements.

% ------------------------------------------------

\item \emph{HD 21120$^\ast$/$o$ Tau}: SIMBAD also shows this star as a spectroscopic binary, though it is not in the WDS. \citet{1957ApJ...125..712J} present an orbit, though no $\Delta m_V$ is indicated. Again, we treat it as single here with the caveat that if the companion is brighter than we expect, it could affect the measurements.

% ------------------------------------------------

\item \emph{HD 22649}: This is a symbiotic system, which is a type of binary system where a white dwarf is accreting material from its red giant partner via Roche-lobe overflow and/or stellar wind \citep{2024ApJ...962..126X}. Presumably the difference in stellar brightness between the two components would be well outside the range of NPOI detection.

% ------------------------------------------------

\item \emph{HD 27371/$\gamma$ Tau}: This Hyades giant star is noted to be resolved using speckle interferometry, and the notes in the WDS say the $\Delta m_V$ is large, though the specific value is not included.

% ------------------------------------------------

\item \emph{HD 29095/58 Per}: Another spectroscopic binary according to SIMBAD, \citet{2020RNAAS...4...12P} identifies the pair as a red supergiant and a B star. \citet{1983MNRAS.204..927H} found $\Delta m_V$ of 2.9 to 3.9 across the wavelengths the NPOI observes, the smallest range of which is just barely detectable using the NPOI. We do not note any particular sign of binarity in the diameter fit.

% ------------------------------------------------

\item \emph{HD 32068$^\ast$/$\zeta$ Aur}: This is an eclipsing binary consisting of a K4 supergiant star and a B5 dwarf with the supergiant passing in front of the dwarf every 972 d \citep{2005Obs...125....1G}\footnote{As an aside, this paper contains a splendid discussion of Miss Maury's spectral classification system that was an alternate to the system developed at Harvard for the Draper Catalog.}. The WDS does not have $\Delta m_V$ listed, and the separation is shown as 0.00 arcsec. \citet{1996ApJ...471..454B} used the Mark III interferometer and the HST Goddard High Resolution Spectrograph to determine masses and radii for the components, and found $\Delta m_V = 2.22$ in the $V$-band. We include the angular diameter measurement in this paper, albeit with caveats, considering we do not see significant signs of the the companion affecting the visibility measurements. 

% ------------------------------------------------

\item \emph{HD 98839/56 UMa}: \citet{2023AandA...670L..14E} classifies this star as a chemically peculiar red giant barium star that has a faint companion. The latter is expected to be a white dwarf or, perhaps, a neutron star, and here we assume that $\Delta m_V$ is too large to detect with the NPOI.

% ------------------------------------------------

\item \emph{HD 139006/$\alpha$ CrB:} This is an eclipsing, double-lined spectroscopic binary \citep{2024AandA...683A.252T}, though the $\Delta m_V$ in the $V$-band is $\sim$4 mag \citep{1986AJ.....91.1428T}.

% ------------------------------------------------

\item \emph{HD 143666/r Her}: This is labeled as a spectroscopic binary on SIMBAD, though \citet{2008Obs...128...21G} says ``the mass function is too small to encourage hope that the secondary object will be detectable at all easily.''

% ------------------------------------------------

\item \emph{HD 164136$^\ast$/94 Her}: SIMBAD lists this star as a long-period variable, and the WDS notes $\Delta m_V = 2.93$ mag with a separation between 0.4 and 0.5 arcsec, and there are no written notes. Considering this is on the edge of the NPOI's detection capabilities, we present the diameter as a single star but note some slight contamination from the secondary may exist.

% ------------------------------------------------

\item \emph{HD 165908/b Her}: The WDS shows $\Delta m_V$ for the AB pair as 3.83 mag, and no $\Delta m_V$ is given for a purported Aa,Ab pair. \citet{2019ApJS..243...32H} used the NPOI to conclude that there is only the AB pair.

% ------------------------------------------------

\item \emph{HD 175535}: Though SIMBAD indicates this is a spectroscopic binary, \citet{2010Obs...130..299G} note that $\Delta m_V$ could be as high as 9 mag for a main-sequence companion, and even larger if it is a white dwarf.

% ------------------------------------------------

\item \emph{HD 181391$^\ast$/f Aql}: One last spectroscopic binary, the WDS entry has no $\Delta m_V$ listed for the Aa,Ab pair. The separation is 0.5 mas, though \citet{2011MNRAS.413.1200R} found the pair to be unresolved. We present it as a single star here but note that if the $\Delta m_V$ is small enough, light from the secondary star could impact the measurement.

% ------------------------------------------------

\item \emph{HD 197345/$\alpha$ Cyg/Deneb}: We comment on this star only because it is one of the brightest and best-known stars in the night sky. Deneb is a luminous blue variable, where the variability period itself varies over time. It is in the WDS, but with $\Delta m_V = 10.45$, so we do not concern ourselves with the companion here.

% ------------------------------------------------

\item \emph{HD 218634/57 Peg}: This is an S star, a class that \citet{1993ApJ...416..769C} define as a peculiar red giant stars, enriched by the $s$-process elements, that fall between more oxygen-rich M stars and carbon-rich C stars. The WDS shows $\Delta m_V = 2.95$ for the Aa,Ab pair, though \citet{2015AJ....150..151H} find $\Delta m = 4.30$ at 692 nm, and \citet{2008MNRAS.389..869E} indicate the most probable multiplicity is 1. We treat it as single here.

% ------------------------------------------------

\item \emph{HD 223047/$\psi$ And}: This is a complicated system with components in the WDS spanning Aa,Ab to AE. No $\Delta m_V$ is provided for the Aa,Ab pair, and the Aa,Ac pair is $\Delta m_V = 2.90$ with separations from 0.2 to 0.3 arcsec. However, the notes in the WDS discuss the possible period of Aa,Ac ($\sim$300 yr) as well as the dynamical stability of the system, speculating that Aa,Ab could have a larger $\Delta m_V$ or if it is a spurious detection. \citet{2019AJ....158..167T} publish $\Delta m$ of 3.5 and 4.3 at 562 and 716 nm, respectively. We treat the star as single.

\end{itemize}

% ------------------------------------------------

\section{Discussion and Conclusion}
\label{sec:disc}

\begin{figure}[h]
\includegraphics[width=0.75\textwidth]{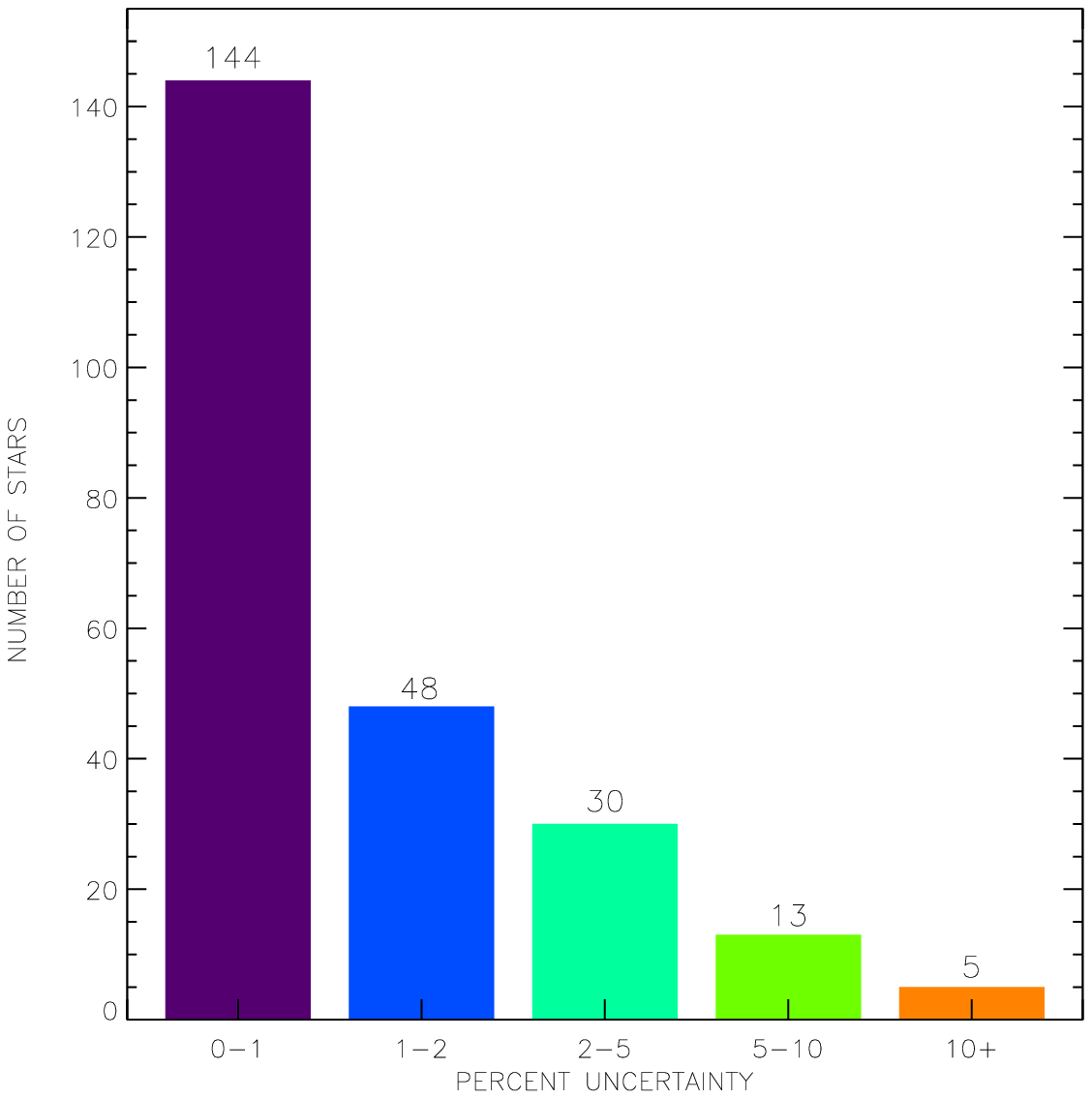}
\caption{The distribution of percent uncertainty for the entire NPOI sample of 241 stars (see Section 6.1 for details).}
  \label{stats}
\end{figure}

We present angular diameters for 145 stars with precisions that range from 0.1$\%$ to 46$\%$, with an average of 1.8$\%$. There are only three stars with errors over 10$\%$, and eight more with errors larger than 5$\%$. When combined with previous NPOI angular diameters, we have built a sample of 241 stars \citep[][see Section 6 and Table 9]{2023AJ....166..268B}. When it comes to determining stellar diameters, much depends on the precision of the measurement, and a 1-2$\%$ precision is key for the star's use as determining abundances and log$g$ \citep{1997IAUS..189..147B}, distinguishing between stellar evolutionary models, as well as when the star is used to evaluate properties of potential exoplanets orbiting it \citep{2011AJ....142..176M, 2017ApJ...845...65N}. Figure \ref{stats} shows the distribution of the percent uncertainties for all 241 stars, and the majority that falls within the 2$\%$ or less categories, 192 stars, is encouraging.

\subsection{Habitable Worlds Observatory Targets}

Thirteen of the stars are identified as promising planet survey target for the Habitable Worlds Observatory by \citet{2023AAS...24111607M}. We compared our angular diameters to those from \citet{2024ApJS..272...30H}, which were derived from SED fits. Some stars agreed very well (five of the 13 stars had percent differences $<$5$\%$), some slightly less so (five more were between 5 and 10$\%$), and others not well at all, differing by up to 27$\%$ (see Table \ref{hwo}). This echoes the need for precise measurements expounded by the \emph{NASA Exoplanet Exploration Program Mission Star List for the Habitable Worlds Observatory}\footnote{https://exoplanetarchive.ipac.caltech.edu/docs/2645$\_$NASA$\_$ExEP$\_$Target$\_$List$\_$HWO$\_$Documentation$\_$2023.pdf} document, which clearly and specifically calls for observations of these targets. 

\subsection{Stellar Oscillators}

Angular diameters, when coupled with stellar oscillation measurements, enable mass estimates for single stars. In these cases, we can use the equations relating the frequency of maximum oscillation power ($\nu_{\rm max}$) to stellar mass ($M$), $R$, and $T_{\rm eff}$ \citep{1995AandA...293...87K, 2021ApJ...919..131H}:

\begin{equation}
\left(\frac{M}{M_\odot}\right) = \left(\frac{\nu_{\rm max}}{\nu_{\rm max,\odot}}\right) \left(\frac{R}{R_\odot}\right)^{2} \left(\frac{T_{\rm eff}}{T_{\rm eff,\odot}}\right)^{\frac{1}{2}},
\end{equation}

where $\nu_{\rm max,\odot}$ = 3090 $\mu$Hz \citep{2011ApJ...743..143H} and $T_{\rm eff,\odot} = 5772$ K \citep{2016AJ....152...41P}. Uncertainties were propagated from published uncertainties in $\nu_{\rm max}$, $R$, and $T_{\rm eff}$. Masses range from 0.9 to 2.4 $M_\odot$, and have a median uncertainty of 12$\%$. Table \ref{osc} shows the results of these calculations for nine of our targets, and uncertainties on the masses range from 5$\%$ to 26$\%$. 

\subsection{Zero Crossers}

\begin{figure}[h]
\includegraphics[width=1.0\textwidth]{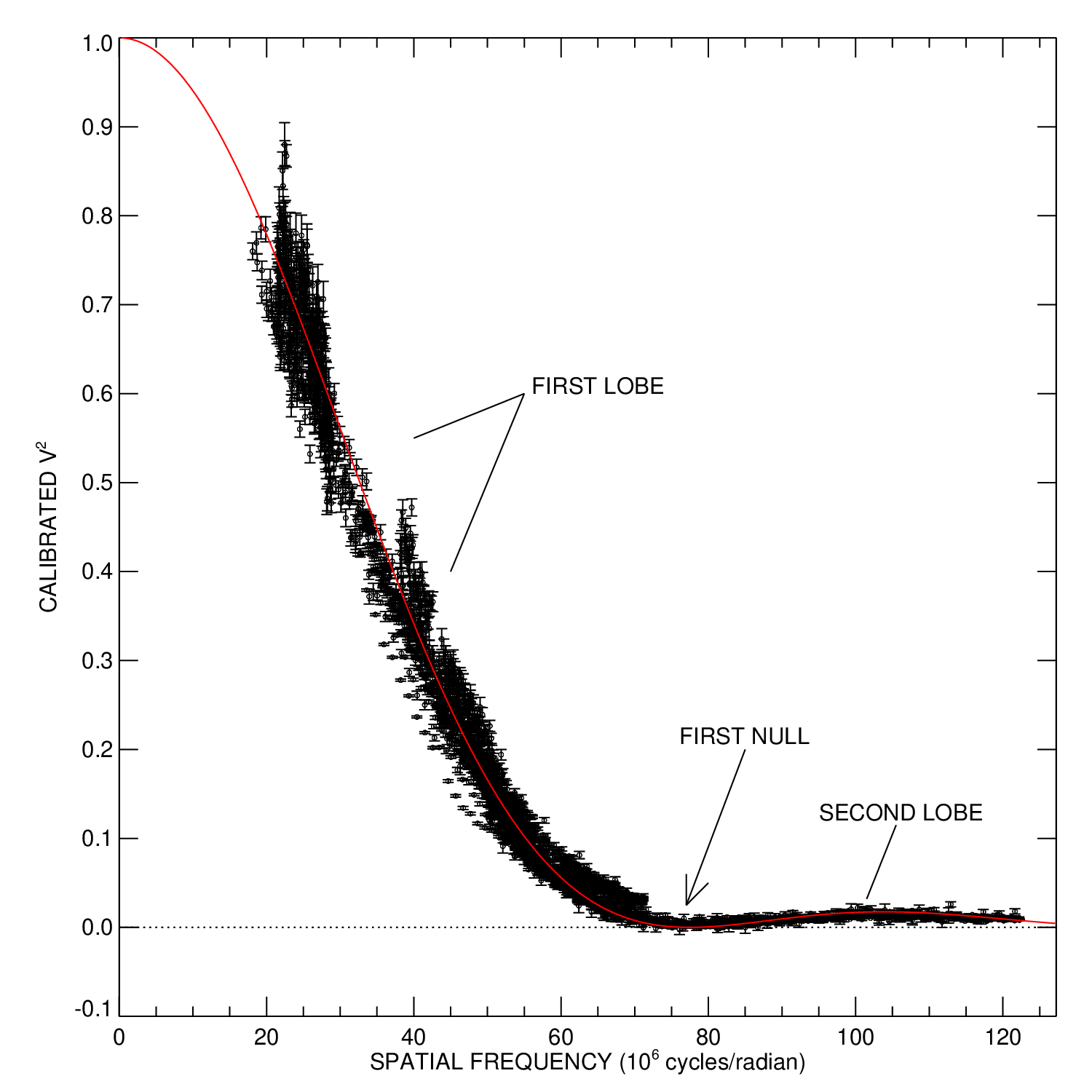}
\caption{An example of a zero-crossing star (HD 187642/$\alpha$ Aql/Altair) with the first lobe, first null, and second lobe indicated.}
  \label{zc_example}
\end{figure}

A final category of stars we would like to discuss are those sufficiently well resolved to reach the ``first null'' in a visibility curve. The majority of interferometric measurements are made on the first lobe of the visibility curve, as can be seen in Figure \ref{diam_fit} and many of the plots in the supplemental figures in the online version of the \emph{Astronomical Journal}. One example of a visibility curve that shows data through the first null, where $V^2$ drops to zero, and onto the second lobe is for HD 187642/$\alpha$ Aql/Altair (Figure \ref{zc_example}). From spatial frequencies of 0 megacycles radian$^{\rm -1}$ to $\sim$75 megacycles radian$^{\rm -1}$, this would be considered the first lobe. The first null would be at that $\sim$75 megacycle radian$^{\rm -1}$ mark, and then the second lobe would be the part between $\sim$75 megacycles radian$^{\rm -1}$ and up. The $V^2$ measurements do not reach the second null in this plot, where $V^2$ returns to 0. For particularly large stars, multiple lobes can be measured \citep[see, e.g., ][]{2009AandA...508..923H}. Measuring visibilities to the first null and beyond are especially important when it comes to directly determining some second-order effects such as limb-darkening and gravity-darknening \citep{2001AandA...377..981W, 2006ApJ...645..664A}. Altair in particular is a gravity-darkened and oblate star, as imaged by \citet{2007Sci...317..342M}.

In our overall effort to publish angular diameters from the NPOI data archive, we found we have 69 zero crossers in the sample presented here, and a further 51 from previous publications (see Table \ref{zc} for a list). Some are very clean at the null and beyond, as is shown in Figure \ref{zc_example}, while others are noisier (e.g., HD 102224/$\chi$ UMa), some just barely touch the $V^2=0$ line (e.g., HD 76294/$\zeta$ Hya), and others even cover much of the second lobe (e.g., HD 61421/$\alpha$ CMi/Procyon). The best candidates for further investigation include HD 3712/$\alpha$ Cas, HD 61241/$\alpha$ CMi/Procyon, HD 62509/$\beta$ Gem/Pollux, HD 89758/34 UMa, HD 96833/$\psi$ UMa, HD 127665/$\rho$ Boo, HD 156283/67 Her, HD 172167/$\alpha$ Lyr/Vega, and HD 187642/$\alpha$ Aql/Altair.\footnote{These are just the candidates from this sample, for the ease of seeing the plots without having to go to different publications.}

% ------------------------------------------------

\begin{acknowledgements}

The Navy Precision Optical Interferometer is funded by the Office of Naval Research and the Oceanographer of the Navy. This research has made use of the SIMBAD database and Vizier catalogue access tool, operated at CDS, Strasbourg, France. This research has made use of the Washington Double Star Catalog maintained at the U.S. Naval Observatory.

\end{acknowledgements}

\clearpage

%%%%%%%%%%%%%%%%%%%%%%%%%%%%%% BASELINES %%%%%%%%%%%%%%%%%%%%%%%%%%%%%%%%%%%%%%%

\begin{deluxetable}{cc||cc}
\tablewidth{0pc}
\tabletypesize{\normalsize}
\tablecaption{NPOI Baselines.\label{baselines}}
\tablehead{\colhead{Telescope Pair} & \colhead{Length (m)} & \colhead{Telescope Pair} & \colhead{Length (m)}}
\startdata
AC-AE & 18.9 & AN-E6 & 45.6 \\
AC-AN & 22.8 & AN-W7 & 66.4 \\
AC-AW & 22.2 & AW-E2 & 29.1  \\
AC-E2 & 7.2  & AW-E3 & 31.9 \\
AC-E6 & 34.4 & AW-E4 & 34.5  \\
AC-W4 & 8.8  & AW-E6 & 53.3 \\
AC-W7 & 51.3 & AW-E7 & 73.5 \\
AE-AN & 34.9 & AW-W4 & 14.0 \\
AE-AW & 37.5 & AW-W7 & 29.5 \\
AE-E2 & 15.9 & E2-E4 & 7.8  \\ 
AE-E6 & 15.9 & E2-W7 & 58.7 \\
AE-E7 & 37.1 & E3-W4 & 18.6 \\
AE-W4 & 26.7 & E6-W4 & 42.5 \\
AE-W7 & 64.2 & E6-W7 & 79.4 \\
AN-AW & 38.2 & E7-W7 & 97.5 \\
AN-E2 & 20.0 &       &      \\
\enddata
\end{deluxetable}

\clearpage

%%%%%%%%%%%%%%%%%%%%%%%%%%%%%% GENERAL PROPERTIES %sec:notes%%%%%%%%%%%%%%%%%%%%%%%%%%%%%%%%%%%%%%

\startlongtable
\begin{deluxetable}{cccccccclcc}
\tablewidth{0pc}
\tabletypesize{\normalsize}
\tablecaption{Sample Star Properties.\label{general_properties}}
\tablehead{
 \colhead{}   & \colhead{}   & \colhead{}    & \colhead{WDS} & \colhead{Other} & \colhead{Spectral} & \colhead{$V$}   & \colhead{Parallax} & \colhead{}    & \colhead{$v \sin i$} & \colhead{Previously}        \\
 \colhead{HD} & \colhead{HR} & \colhead{FK5} & \colhead{ID}  & \colhead{Name}  & \colhead{Type}     & \colhead{(mag)} & \colhead{(mas)}    & \colhead{Ref} & \colhead{(km/s)} & \colhead{Published?} }
\startdata
432    & 21   & 2    & 00092+5909 & $\beta$ Cas    & F2 III         & 2.27 & 59.58$\pm$0.38  & vL07    & 70  & Y \\
3712   & 168  & 21   & 00405+5632 & $\alpha$ Cas   & K0 III         & 2.23 & 14.09$\pm$0.50  & Gaia22  & 21  & Y \\
5394   & 264  & 32   & 00567+6043 & $\gamma$ Cas   & B0.5 IV        & 2.48 & 5.94$\pm$0.12   & vL07    & 300 & N \\
6186   & 294  & 36   & 01029+0753 & $\epsilon$ Psc & G9 III         & 4.27 & 17.81$\pm$0.18  & Gaia22  & 17  & Y \\
7087   & 351  & 1032 & -          & $\chi$ Psc     & G8.5 III       & 4.66 & 8.06$\pm$0.12   & Gaia22  & 19 & N \\
8890   & 424  & 907  & 02318+8916 & $\alpha$ UMi   & F8 I           & 2.01 & 7.54$\pm$0.11   & vL07    & 17 & N \\
9270   & 437  & 50   & 01315+1521 & $\eta$ Psc     & G7 III         & 3.62 & 8.73$\pm$0.92   & Gaia22  & 19 & N \\
9408   & 442  & -    & -          & $\chi$ Cas     & G9 III         & 4.70 & 16.26$\pm$0.10  & Gaia22  & 19 & N \\
\enddata
\tablecomments{Spectral types are from SIMBAD, $V$ magnitudes are from the Catalogue of Homogeneous Means in the UBV System \citep{Mermilliod}, parallaxes are from the sources listed in Table \ref{table_refs}, and $v \sin i$ is from the Catalog of Bright Stars \citep{1964cbs..book.....H}. The final column indicates if the stellar diameter has been previously published by our group.}
\end{deluxetable}

\clearpage

%%%%%%%%%%%%%%%%%%%%%%%%%%%%%% TABLE REFERENCES %%%%%%%%%%%%%%%%%%%%%%%%%%%%%%%%%%%%%%%

\begin{deluxetable}{ll||ll}
\tablewidth{0pc}
\tabletypesize{\normalsize}
\tablecaption{Table References.\label{table_refs}}
\tablehead{\colhead{Abbr} & \colhead{Reference} & \colhead{Abbr} & \colhead{Reference}}
\startdata
AF12   & \citet{2012AstL...38..331A}   & K19   & \citet{2019AandA...623A..72K} \\
A23    & \citet{2023ApJS..266....6A}   & KAT22 & \citet{2022AandA...657A...7K} \\
A96    & \citet{1996AandAS..117..227A} & KS06  & \citet{2006MNRAS.371.1793K} \\
APL99  & \citet{1999AandA...352..555A} & LB03  & \citet{2003AandA...402..433L} \\
B02    & \citet{2002AandA...393..183B} & M11   & \citet{2011MNRAS.414.1227M} \\
B96    & \citet{1996AandAS..118..481B} & MZW17 & \citet{2017MNRAS.471..770M} \\
C04    & \citet{2004AJ....127.1227C}   & N80   & \citet{1980BICDS..19...61N} \\
C13    & \citet{2013AandA...549A.129C} & OD06  & \citet{2006ApJ...647..312O} \\
\enddata
\end{deluxetable}

\clearpage

%%%%%%%%%%%%%%%%%%%%%%%%%%%%%% OBSERVING LOG %%%%%%%%%%%%%%%%%%%%%%%%%%%%%%%%%%%%%%%

\begin{deluxetable}{ccllcc}
\tablewidth{0pc}
\tabletypesize{\normalsize}
\tablecaption{Observing Log.\label{observations}}
\tablehead{
 \colhead{Target} & \colhead{Calibrator} & \colhead{Date} & \colhead{Baselines} & \colhead{$\#$}  & \colhead{$\#$}  \\
 \colhead{HD}     & \colhead{HD}         & \colhead{(UT)} & \colhead{Used}      & \colhead{Scans} & \colhead{Data Points}  }
\startdata
432    & 886    & 1997 Sep 24 & AC-AE, AC-AW, AE-AW                      & 12 & 600 \\
       & 3360   & 2005 Sep 29 & AW-E6                                    & 10 & 150 \\
       &        & 2005 Sep 30 & AC-AE, AC-AN, AC-AW, AC-E6, AE-AN, AW-E6 & 10 & 600 \\
       &        & 2005 Oct 3  & AC-AN, AE-AN                             & 4  &  80 \\
       &        & 2005 Oct 6  & AC-AW, AC-E6, AW-E6                      & 3  & 120 \\
       &        & 2005 Oct 7  & AC-AN, AC-E6, AE-AN, AW-E6               & 8  & 360 \\
       &        & 2005 Oct 8  & AW-E6                                    & 6  &  89 \\
       &        & 2005 Oct 11 & AW-E6                                    & 10 & 150 \\
\enddata
\tablecomments{See Table \ref{baselines} for the baseline lengths. ``$\#$ Scans'' is the number of scans per baseline, and ``$\#$ Data Points'' is the number of calibrated $V^2$ measurements used in the angular diameter fit. This table shows the information for two stars as an example; the full table is available on the electronic version of the \emph{Astronomical Journal}.}
\end{deluxetable}

\clearpage

%%%%%%%%%%%%%%%%%%%%%%%%%%%%% CALIBRATORS %%%%%%%%%%%%%%%%%%%%%%%%%%%%%%%%%%%

\startlongtable 
\begin{longrotatetable}
\begin{deluxetable}{cccccccccccclclc}
\tablewidth{0pc}
\tabletypesize{\small}
\tablecaption{Calibrator Stars' SED Inputs and Resulting Angular Diameters. \label{calibrators}}
\tablehead{
 \colhead{ }  & \colhead{Spec} & \colhead{$U$}   & \colhead{$B$}   & \colhead{$V$}   & \colhead{$R$}   & \colhead{$I$}   & \colhead{$J$}   & \colhead{$H$}   & \colhead{$K$}   & \colhead{$T_{\rm eff}$} & \colhead{log $g$}     & \colhead{}    & \colhead{}         & \colhead{}    & \colhead{$\theta_{\rm est}$} \\
 \colhead{HD} & \colhead{Type} & \colhead{(mag)} & \colhead{(mag)} & \colhead{(mag)} & \colhead{(mag)} & \colhead{(mag)} & \colhead{(mag)} & \colhead{(mag)} & \colhead{(mag)} & \colhead{(K)}           & \colhead{(cm s$^{-2}$)} & \colhead{Ref} & \colhead{$E(B-V)$} & \colhead{Ref} & \colhead{(mas)}   }
\startdata
886    & B2 IV       & 1.75 & 2.61 & 2.83 & 2.88 & 3.06 & 3.50 & 3.64 & 3.77 & 21944 & 3.93 & P07   & 0.02 & SB06  & 0.45$\pm$0.02 \\
2905   & B0 III      & 3.50 & 4.30 & 4.16 & 4.12 & 4.09 & 4.14 & 4.15 & 4.01 & 21500 & 2.45 & Sou16 & 0.33 & S85   & 0.36$\pm$0.02 \\
3360   & B2 IV       & 2.62 & 3.47 & 3.66 & 3.74 & 3.92 & 4.14 & 4.25 & 4.25 & 22180 & 3.92 & V04   & 0.03 & Z09   & 0.32$\pm$0.02 \\
5448   & A6 V        & 4.14 & 3.99 & 3.86 & 3.78 & 3.71 & 3.62 & 3.65 & 3.64 & 8128  & 3.78 & APL99 & 0.00 & N80   & 0.69$\pm$0.03 \\
6961   & A7 V        & 4.63 & 4.51 & 4.34 & 4.25 & 4.16 & 4.45 & 4.28 & 4.13 & 7762  & 3.80 & APL99 & 0.00 & C04   & 0.58$\pm$0.03 \\
7964   & A3 V        & 4.88 & 4.77 & 4.75 & 4.73 & 4.73 & 4.85 & 4.60 & 4.54 & 8913  & 3.65 & APL99 & 0.04 & N80   & 0.42$\pm$0.02 \\
11415  & B3 V        & 2.62 & 3.22 & 3.37 & 3.40 & 3.53 & 3.86 & 3.93 & 3.96 & 14250 & 3.38 & HGM10 & 0.05 & F92   & 0.50$\pm$0.03 \\
12216  & A2 V        & 4.00 & 3.96 & 3.97 & 3.93 & 3.94 & 3.89 & 3.90 & 3.92 & 9333  & 3.94 & APL99 & 0.01 & N80   & 0.56$\pm$0.03 \\
\enddata
\tablecomments{Spectral types are from SIMBAD; $UBV$ values are from the Catalogue of Homogeneous Means in the UBV System \citep{Mermilliod}; $RI$ values are from the The USNO-B Catalog \citep{2003AJ....125..984M}; $JHK$ values are from the 2MASS All-Sky Catalog of Point Sources \citep{2003yCat.2246....0C}; $T_{\rm eff}$, log $g$, and $E(B-V)$ values are from the sources listed in Table \ref{table_refs}. $\theta_{\rm est}$ is the estimated angular diameter calculated using the method described in Section \ref{sec:obsvis}.}
\end{deluxetable}
\end{longrotatetable}

%\clearpage

%%%%%%%%%%%%%%%%%%%%%% INTERFEROMETRIC RESULTS %%%%%%%%%%%%%%%%%%%%%%%%%%%%%%%%%%%%%%%%%%%%%%%
\startlongtable
\begin{longrotatetable}
\begin{deluxetable}{cccclclccccccc}
\tablewidth{0pc}
\tabletypesize{\normalsize}
\tablecaption{Interferometric Results. \label{inf_results}}

\tablehead{\colhead{Target} & \colhead{$\theta_{\rm UD}$} & \colhead{$T_{\rm eff}$} & \colhead{log $g$}        & \colhead{ }   & \colhead{ }      & \colhead{ }   & \colhead{Initial}             & \colhead{$\theta_{\rm LD,initial}$} & \colhead{Final}               & \colhead{$\theta_{\rm LD,final}$} & \colhead{$\sigma_{\rm LD}$} & \colhead{Max SF} & \colhead{$\#$} \\ 
           \colhead{HD}     & \colhead{(mas)}             & \colhead{(K)}           & \colhead{(cm s$^{-2}$)}  & \colhead{Ref} & \colhead{[Fe/H]} & \colhead{Ref} & \colhead{$\mu_{\rm \lambda}$} & \colhead{(mas)}             & \colhead{$\mu_{\rm \lambda}$} & \colhead{(mas)}             & \colhead{($\%$)} & \colhead{(10$^6$cycles rad$^{\rm -1}$)}   & \colhead{pts}            }
\startdata
432    & 1.994$\pm$0.008 & 6894  & 3.47 & MZW17   &  0.03 & AF12  & 0.48 & 2.082$\pm$0.009 & 0.49 & 2.085$\pm$0.008 & 0.4 & 94.9  & 2525 \\
3712   & 5.104$\pm$0.016 & 4690  & 1.19 & MZW17   & -0.10 & AF12  & 0.67 & 5.527$\pm$0.018 & 0.67 & 5.527$\pm$0.018 & 0.3 & 71.1  & 1337 \\
5394   & 1.200$\pm$0.011 & 35210 & 4.18 & Z16     &  0.02 & AF12  & 0.24 & 1.223$\pm$0.011 & 0.40 & 1.241$\pm$0.011 & 0.9 & 124.6 & 3675 \\
6186   & 1.771$\pm$0.013 & 4828  & 2.35 & MZW17   & -0.29 & AF12  & 0.64 & 1.883$\pm$0.013 & 0.67 & 1.890$\pm$0.014 & 0.7 & 98.0  & 3621 \\
7087   & 1.118$\pm$0.067 & 4812  & 2.13 & Gaia22  & -0.11 & AF12  & 0.67 & 1.190$\pm$0.071 & 0.59 & 1.181$\pm$0.071 & 6.0 & 71.1  & 1018 \\
8890   & 3.110$\pm$0.005 & 5904  & 1.59 & C21     &  0.13 & AF12  & 0.53 & 3.276$\pm$0.006 & 0.55 & 3.284$\pm$0.006 & 0.2 & 71.1  & 4886 \\
9270   & 2.187$\pm$0.026 & 4898  & 2.26 & APL99   & -0.06 & AF12  & 0.66 & 2.330$\pm$0.027 & 0.66 & 2.330$\pm$0.027 & 1.2 & 71.0  & 1453 \\
9408   & 1.464$\pm$0.045 & 4843  & 2.34 & MZW17   & -0.23 & AF12  & 0.67 & 1.559$\pm$0.048 & 0.64 & 1.553$\pm$0.048 & 3.1 & 71.1  & 1331 \\
\enddata 
\tablecomments{$^\ast$See Section 5 for discussion about potential sources of bias for this star. \\
The initial $\mu_{\lambda}$ is based on the $T_{\rm eff}$ and log $g$ referenced in the fifth column, and the final $\mu_{\lambda}$ is based on the new $T_{\rm eff}$ determination (see Section \ref{sec:randt} for more details). The $T_{\rm eff}$, log $g$, and [Fe/H] are from the sources listed in Table \ref{table_refs}. Max SF is the maximum spatial frequency for that star's diameter measurement, and $\#$ pts is the number of data points used for the angular diameter fit.
} 
\end{deluxetable}
\end{longrotatetable}

\clearpage

%%%%%%%%%%%%%%%%%%%%%% DERIVED RESULTS %%%%%%%%%%%%%%%%%%%%%%%%%%%%%%%%%%%%%%%%%%%%%%%

\startlongtable 
\begin{deluxetable}{cccclccc}
\tablewidth{0pc}
\tabletypesize{\normalsize}
\tablecaption{Derived Stellar Parameters. \label{derived_results}}

\tablehead{\colhead{Target} & \colhead{$R$}         & \colhead{$\sigma_R$} & \colhead{$L$}       & \colhead{}    & \colhead{$F_{\rm BOL}$}                      & \colhead{$T_{\rm eff}$} & \colhead{$\sigma_T$}         \\ 
           \colhead{HD}     & \colhead{($R_\odot$)} & \colhead{($\%$)}     & \colhead{$L_\odot$} & \colhead{Ref} & \colhead{(10$^{-6}$ erg s$^{-1}$ cm$^{-2}$)} & \colhead{(K)}           & \colhead{($\%$)}   }
\startdata
432    &   3.76$\pm$0.03             &  0.7 &    25.8$\pm$0.1   & MZW17 &  292.7$\pm$2.0  & 6705$\pm$17  & 0.3 \\
3712   &  42.15$^{+1.45}_{-1.55}$    &  3.6 &   733.5$\pm$0.1   & MZW17 &  465.6$\pm$16.5 & 4625$\pm$42  & 0.9 \\
5394   &  22.45$^{+0.49}_{-0.50}$    &  2.2 &  4545.7$\pm$227   & AF12  &  512.8$\pm$27.7 & 9998$\pm$142 & 1.4 \\
6186   &  11.41$\pm$0.14             &  1.3 &    62.2$\pm$0.1   & MZW17 &   63.0$\pm$0.6  & 4797$\pm$22  & 0.5 \\
7087   &  15.75$\pm$0.97             &  6.2 &   192.6$\pm$0.1   & MZW17 &   40.0$\pm$0.6  & 5416$\pm$164 & 3.0 \\
8890   &  46.81$^{+0.68}_{-0.70}$    &  1.5 &  2219.7$\pm$0.0   & MZW17 &  403.4$\pm$5.9  & 5789$\pm$22  & 0.4 \\
9270   &  28.67$^{+2.75}_{-3.40}$    & 10.6 &   408.0$\pm$0.1   & MZW17 &   99.5$\pm$10.5 & 4843$\pm$131 & 2.7 \\
9408   &  10.27$\pm$0.32             &  3.1 &    58.0$\pm$0.1   & MZW17 &   49.0$\pm$0.3  & 4970$\pm$77  & 1.6 \\
\enddata
\tablecomments{See Section \ref{sec:randt} for a description of how these parameters were derived, and $L$ is from the sources listed in Table \ref{table_refs}.}
\end{deluxetable}

%%%%%%%%%%%%%%%%%%%%%%%%%%%%%% WDS TARGETS %%%%%%%%%%%%%%%%%%%%%%%%%%%%%%%%%%%%%%%

\begin{deluxetable}{ccccc}
\tablewidth{0pc}
\tabletypesize{\normalsize}
\tablecaption{Stars in the WDS.\label{wds}}
\tablehead{
 \colhead{Target} & \colhead{ } & \colhead{$\Delta m_V$} & \colhead{$a$} & \colhead{Needs} \\
 \colhead{HD} & \colhead{Components} & \colhead{(mag)} & \colhead{(arcsec)} & \colhead{Discussion?} }
\startdata
5394   & AB    & 8.70  & 2.15   & N \\
8890   & Aa,Ab & 2.00  & 0.15   & Y \\
9270   & -     & 3.68  & 0.80   & N \\
11353  & Aa,Ab & -     & 0.10   & Y \\
20630  & AB    & 3.94  & 267.50 & N \\
25604  & AC    & 5.55  & 136.80 & N \\
27371  & Aa,Ab & -     & 0.40   & Y \\
27697  & Aa,Ab & 5.60  & 0.40   & N \\
\enddata
\tablecomments{Some of these systems are large and complex, so we only consider the inner-most binary star pair because the other stars are well outside our range of detection of 0.86 arcsec and $\Delta m_V$ of 3.0. The exceptions to this are HD 165908 and HD 223047, so they have multiple lines in this table. See Section \ref{sec:notes} for more discussion of the stars that require it.}
\end{deluxetable}

\clearpage

%%%%%%%%%%%%%%%%%%%%%%%%%%%%%% HWO TARGETS %%%%%%%%%%%%%%%%%%%%%%%%%%%%%%%%%%%%%%%

\begin{deluxetable}{cccc}
\tablewidth{0pc}
\tabletypesize{\normalsize}
\tablecaption{Promising HWO Targets.\label{hwo}}
\tablehead{
 \colhead{Target} & \colhead{$R_{\rm NPOI}$} & \colhead{$R_{\rm HWO}$} & \colhead{$\%$} \\
 \colhead{HD}     & \colhead{($R_\odot$)} & \colhead{($R_\odot$)}      & \colhead{Diff}     }
\startdata
9826   & 1.083$\pm$0.018 & 1.110 & 2 \\
10700  & 2.072$\pm$0.010 & 2.103 & 1 \\
19373  & 1.017$\pm$0.031 & 1.235 & 21 \\
20630  & 0.949$\pm$0.049 & 0.949 & 0 \\
23249  & 2.088$\pm$0.143 & 2.297 & 10 \\
102870 & 1.825$\pm$0.054 & 1.429 & 22 \\
109358 & 1.120$\pm$0.019 & 1.136 & 1 \\
114710 & 1.159$\pm$0.021 & 1.092 & 6 \\
126660 & 1.160$\pm$0.032 & 1.086 & 6 \\
128167 & 0.714$\pm$0.205 & 0.771 & 8 \\
141004 & 0.982$\pm$0.056 & 1.058 & 8 \\
142373 & 0.802$\pm$0.036 & 1.019 & 27 \\
165341 & 1.590$\pm$0.102 & 1.574 & 1 \\
\enddata
\tablecomments{$R_{\rm HWO}$ is the radius predicted by the \emph{NASA ExEP Mission Star List for HWO} document linked in Section 6.2.}
\end{deluxetable}

\clearpage

%%%%%%%%%%%%%%%%%%%%%%%%%%%%%% OSCILLATORS %%%%%%%%%%%%%%%%%%%%%%%%%%%%%%%%%%%%%%%

\begin{deluxetable}{cccccc}
\tablewidth{0pc}
\tabletypesize{\normalsize}
\tablecaption{Stellar Oscillators.\label{osc}}
\tablehead{
 \colhead{Target} & \colhead{$R$}       & \colhead{$T_{\rm eff}$} & \colhead{$\nu_{\rm max}$} & \colhead{$M$}  & \colhead{$\sigma_M$}  \\
 \colhead{HD}     & \colhead{($R_\odot$)} & \colhead{(K)}           & \colhead{($\mu$Hz)}     & \colhead{($M_\odot$)}  & \colhead{$\%$}  }
\startdata
9408   & 10.27$\pm$0.32 & 4970$\pm$77  &  28.6$\pm$2.5  & 0.91$\pm$0.10 & 11 \\
23249  &  2.03$\pm$0.04 & 6676$\pm$322 & 676.0$\pm$5.0  & 0.97$\pm$0.05 & 5 \\
82741  &  8.78$\pm$0.61 & 5256$\pm$181 &  66.4$\pm$12.2 & 1.58$\pm$0.37 & 23 \\
137759 & 11.79$\pm$0.06 & 4621$\pm$8   &  39.5$\pm$7.8  & 1.59$\pm$0.31 & 20 \\
143666 & 10.87$\pm$0.56 & 4776$\pm$75  &  39.7$\pm$2.6  & 1.38$\pm$0.17 & 12 \\
147677 &  6.46$\pm$0.82 & 5642$\pm$358 &  99.4$\pm$4.0  & 1.33$\pm$0.34 & 26 \\
163588 & 11.52$\pm$0.06 & 4504$\pm$9   &  41.3$\pm$6.3  & 1.57$\pm$0.24 & 15 \\
176524 & 22.28$\pm$0.68 & 4217$\pm$40  &  17.1$\pm$1.6  & 2.35$\pm$0.26 & 11 \\
180610 &  8.87$\pm$0.17 & 4868$\pm$124 &  53.3$\pm$2.5  & 1.25$\pm$0.08 & 6 \\
181984 & 11.24$\pm$0.25 & 4339$\pm$39  &  34.2$\pm$2.5  & 1.21$\pm$0.10 & 9 \\
206952 & 10.80$\pm$0.11 & 4926$\pm$21  &  48.9$\pm$3.0  & 1.71$\pm$0.11 & 6 \\
212496 & 10.72$\pm$0.10 & 4717$\pm$17  &  36.9$\pm$4.6  & 1.24$\pm$0.16 & 13 \\
218031 & 11.64$\pm$0.34 & 4345$\pm$63  &  28.7$\pm$1.7  & 1.09$\pm$0.09 & 8 \\
218452 & 23.36$\pm$0.59 & 4332$\pm$291 &   8.5$\pm$1.1  & 1.30$\pm$0.19 & 14 \\
\enddata
\tablecomments{$\nu_{\rm max}$ is from \citet{2021ApJ...919..131H} for all the stars except HD 23249, which is from \citet{2024AandA...683A.161C}.}
\end{deluxetable}

%%%%%%%%%%%%%%%%%%%%%%%%%%%%%% ZC %%%%%%%%%%%%%%%%%%%%%%%%%%%%%%%%%%%%%%%

\begin{deluxetable}{c|c|c|c|c}
\tablewidth{0pc}
\tabletypesize{\normalsize}
\tablecaption{Zero-Crossing Stars.\label{zc}}
\tablehead{
 \colhead{HD} & \colhead{HD} & \colhead{HD} & \colhead{HD} & \colhead{HD}  }
\startdata
\hline									
\multicolumn{5}{c}{ZC stars from this paper} \\									
\hline	
3712  & 81797  & 113226 & 154143 & 188512 \\
12929 & 84441  & 119228 & 156283 & 189319 \\
22649 & 86663  & 120933 & 159561 & 192909 \\
32068 & 87837  & 121370 & 161096 & 198026 \\
48329 & 89758  & 123657 & 161797 & 200905 \\
61421 & 92523  & 127665 & 163588 & 203504 \\
61935 & 94264  & 129989 & 163993 & 206778 \\
62345 & 95689  & 133165 & 169414 & 213311 \\
62509 & 96833  & 133208 & 172167 & 214665 \\
69267 & 97778  & 135722 & 173764 & 216131 \\
73108 & 98262  & 137759 & 176678 & 218329 \\
74442 & 100029 & 145713 & 181276 & 222107 \\
76294 & 102212 & 148856 & 186791 & 222404 \\
80493 & 102224 & 150680 & 187642 &  \\
\hline									
\multicolumn{5}{c}{ZC stars from previous papers} \\									
\hline									
1013  & 31964 & 82308  & 140573 & 204867 \\
1522  & 34085 & 83618  & 141477 & 208816 \\
3627  & 38944 & 108907 & 145713 & 210745 \\
4656  & 42995 & 112300 & 146051 & 211388 \\
5112  & 43232 & 117675 & 153210 & 215182 \\
6805  & 50778 & 120477 & 163917 & 219215 \\
8512  & 54719 & 121130 & 164058 & 219615 \\
9927  & 58207 & 129712 & 168723 &  \\
17709 & 60522 & 131873 & 175588 &  \\
20644 & 70272 & 132813 & 183439 &  \\
22049 & 73108 & 136726 & 200905 &  \\
\enddata
\end{deluxetable}

\clearpage

%%%%%%%%%%%%%%%%%%%%%%% FIGURES %%%%%%%%%%%%%%%%%

\clearpage

\clearpage

\begin{figure}[h]
\includegraphics[width=1.0\textwidth]{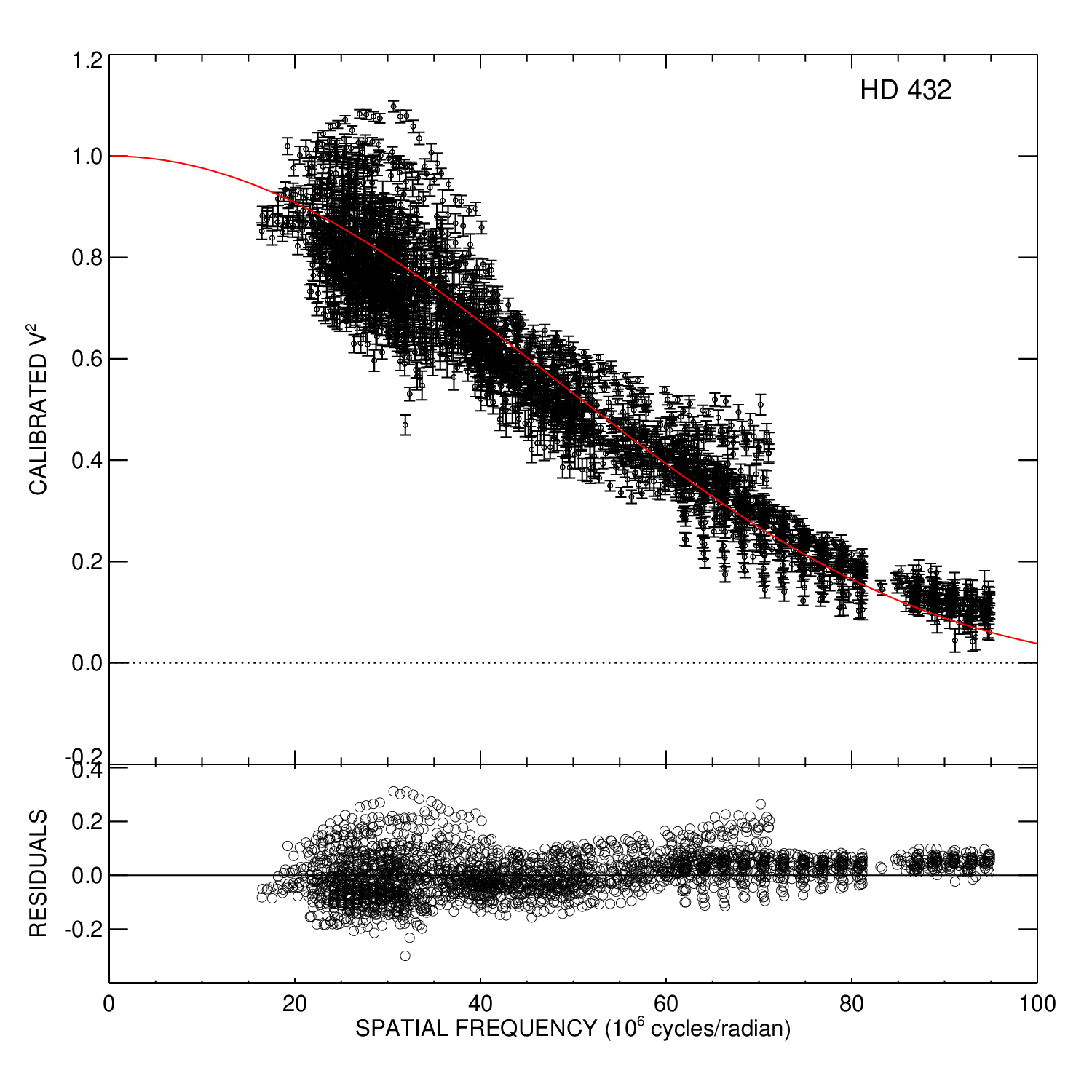}
\caption{Top panel: The $\theta_{\rm LD}$ fit for HD 432/$\beta$ Cas. The solid red line represents the visibility curve for the best fit $\theta_{\rm LD}$, the points are the calibrated visibilities, and the vertical lines are the measurement uncertainties. Bottom panel: The residuals show the difference between the measurement and the visibility curve for each data point. The plots for the remaining stars are available on the electronic version of the \emph{Astronomical Journal}.}
  \label{diam_fit}
\end{figure}

\clearpage

\clearpage

\end{document}